\documentclass[iop]{emulateapj}
\usepackage{color}

\begin{document}

\title{Understanding The Effects Of Stellar Multiplicity On The Derived Planet 
Radii From Transit Surveys: Implications for Kepler, K2, and TESS}

\author{David R.~Ciardi\altaffilmark{1},
Charles A.~Beichman\altaffilmark{1},
Elliott P.~Horch\altaffilmark{2},
Steve B.~Howell\altaffilmark{3}
}
                             
\altaffiltext{1}{NASA Exoplanet Science Institute/Caltech Pasadena, CA USA}
\altaffiltext{2}{Department of Physics, Southern Connecticut State University, 
New Haven, CT, USA}
\altaffiltext{3}{NASA Ames Research Center, Mountain View, CA, USA}

\email{ciardi@ipac.caltech.edu}

%\submitted{Submitted: \today}

%\accepted{11 March 2015, \apj}

\slugcomment{Accepted for publication in The Astrophysical Journal}

\begin{abstract} 
We present a study on the effect of undetected stellar companions on the derived 
planetary radii for the Kepler Objects of Interest (KOIs).  The current 
production of the KOI list assumes that the each KOI is a single star. Not 
accounting for stellar multiplicity statistically biases the planets towards 
smaller radii. The bias towards smaller radii depends on the properties of the 
companion stars and whether the planets orbit the primary or the companion 
stars.  Defining a planetary radius correction factor $X_R$, we find that if the 
KOIs are assumed to be single, then, {\it on average}, the planetary radii may 
be underestimated by a factor of $\langle X_R \rangle \approx 1.5$.  If typical 
radial velocity and high resolution imaging observations are performed and no 
companions are detected, this factor reduces to $\langle X_R \rangle \approx 
1.2$.   The correction factor $\langle X_R \rangle$ is dependent upon the 
primary star properties and ranges from $\langle X_R \rangle \approx 1.6$ for A 
and F stars to $\langle X_R \rangle \approx 1.2$ for K and M stars. For missions 
like K2 and TESS where the stars may be closer than the stars in the Kepler 
target sample, observational vetting (primary imaging) reduces the radius 
correction factor to $\langle X_R \rangle \approx 1.1$.  Finally, we show that 
if the stellar multiplicity rates are not accounted for correctly, occurrence 
rate calculations for Earth-sized planets may overestimate the frequency of 
small planets by as much as $15-20$\%.
\end{abstract}


\section{Introduction}\label{sec-intro}

The Kepler Mission \cite{borucki10}, with the discovery of over 4100 planetary 
candidates in 3200 systems, has spawned a revolution in our understanding of 
planet occurrence rates around stars of all types.  One of Kepler's profound 
discoveries is that small planets ($R_p \lesssim 3 R_\oplus$) are nearly 
ubiquitous \citep[e.g.,][]{howard12,dc13,fressin13,petigura13,batalha14} and, in 
particular, some of the most common planets have sizes between Earth-sized and 
Neptune-sized -- a planet type not found in our own solar system.  Indeed, it is 
within this group of super-Earths to mini-Neptunes that there is a transition 
from ``rocky'' planets to ``non-rocky planets''; the transition is near a planet 
radius of $1.6 R_\oplus$ and is very sharp -- occurring within $\approx 0.2 
R_\oplus$ of this transition radius \citep{marcy14,rogers14}.

Unless an intra-system comparison of planetary radii is performed where only the 
relative planetary sizes are important \citep{ciardi13}, having accurate (as 
well as precise) planetary radii is crucial to our comprehension of the 
distribution of planetary structures.  In particular, understanding the radii of 
the planets to within $\sim 20\%$ is necessary if we are to understand the 
relative occurrence rates of ``rocky'' to ``non-rocky'' planets, and the 
relationship between radius, mass, and bulk density..  While there has been a 
systematic follow-up observation program to obtain spectroscopy and high 
resolution imaging, only approximately half of the Kepler candidate stars have 
been observed (mostly as a result of the brightness distribution of the 
candidate stars). Those stars that have been observed have been done mostly to 
eliminate false positives, to determine the stellar parameters of host stars, 
and to search for nearby stars that may be blended in the Kepler photometric 
apertures.

Stars that are identified as possible binary or triple stars are noted on the 
Kepler Community Follow-Up Observation Program 
website\footnote{https://cfop.ipac.caltech.edu/}, and are often handled in 
individual papers \citep[e.g.,][]{star14,everett14}. The false positive 
assessment of an KOI (or all of the KOIs) can take into account the likelihood 
of stellar companions \citep[e.g.,][]{mj11,morton12}, and a false positive 
probability will likely be included in future KOI lists.  But presently, the 
current production of the planetary candidate KOI list and the associated 
parameters are derived assuming that {\it all} of the KOI host stars are single. 
That is, the Kepler pipeline treats each Kepler candidate host star as a single 
star \citep[e.g.,][]{batalha11,burke14,mullally15}. Thus, statistical studies 
based upon the Kepler candidate lists are also assuming that all the stars in 
the sample set are single stars.

The exact fraction of multiple stars in the Kepler candidate list is not yet 
determined, but it is certainly not zero.  Recent work suggests that a 
non-negligible fraction ($\sim30-40\%$) of the Kepler host stars may be multiple 
stars \citep{adams12,adams13,law14,dressing14,horch14}, although other work may 
indicate that (giant) planet formation may be suppressed in multiple star 
systems \citep{wang14a,wang14b}.   The presence of a stellar companion does not 
necessarily invalidate a planetary candidate, but it does change the observed 
transit depths and, as a result, the planetary radii. Thus, assuming all of the 
stars in the Kepler candidate list are single can introduce a systematic 
uncertainty into the planetary radii and occurrence rate distributions.  This 
has already been discussed for the occurrence rate of hot Jupiters in the Kepler 
sample where it was found that $\sim 13\%$ of hot Jupiters were classified as 
smaller planets because of the unaccounted effects of transit dilution from 
stellar companions \citep{wang14c}.

In this paper, we explore the effects of undetected stellar gravitationally 
bound companions on the observed transit depths and the resulting derived 
planetary radii for the entire Kepler candidate sample.  We do not consider the 
dilution effects of line-of-sight background stars, rather only potential bound 
companions, as companions within $1^{\prime\prime}$ are most likely bound 
companions \cite[e.g.,][]{horch14,gilliland14}, and most stars beyond 
$1^{\prime\prime}$ are either in the Kepler Input Catalog \citep{brown11} or in 
the UKIRT survey of the Kepler field and, thus, are already accounted for with 
regards to flux dilution in the Kepler project transit fitting pipeline. Within 
1\arcsec, the density of blended background stars is fairly low, ranging between 
$0.001 - 0.007$ stars/$\square\arcsec$ \citep{lb14}.  Thus, within a radius of 
1\arcsec, we expect to find a blended background (line-of-sight) star only $0.3 
- 2$\% of the time.  Therefore, the primary contaminant within 1\arcsec\ of the 
host stars are bound companions.

We present here probabilistic uncertainties of the planetary radii based upon 
expected stellar multiplicity rates and stellar companion sizes. We show that, 
in the absence of any spectroscopic or high resolution imaging observations to 
vet companions, the observed planetary radii will be systematically too small. 
However, if a candidate host star is observed with high resolution imaging (HRI) 
or with radial velocity (RV) spectroscopy to screen the star for companions, the 
underestimate of the true planet radius is significantly reduced.  While imaging 
and radial velocity vetting is effective for the Kepler candidate host stars, it 
will be even more effective for the K2 and TESS candidates which will be, on 
average, 10 times closer than the Kepler candidate host stars.

\section{Effects of Companions on Planet Radii}\label{sec-effects}

The planetary radii are not directly observed; rather, the transit depth is the 
observable which is then related to the planet size. The observed depth 
($\delta_o$) of a planetary transit is defined as the fractional difference in 
the measured out-of-transit flux ($F_{total}$) and the measured in-transit flux 
$(F_{transit})$:
\begin{equation}
\delta_o = \frac{F_{total} - F_{transit}} {F_{total}}.\label{eq-single-flux}
\end{equation}
If there are $N$ stars within a system, then the total out-of-transit flux in 
the system is given by 
\begin{equation}
F_{total} = \sum_{i=1}^{N}F_i\label{eq-total-flux},
\end{equation}
and if the planet transits the $t^{th}$ star in the system, then
the in--transit flux can be defined as
\begin{equation}
F_{transit} = F_{total} - F_t\left(R_p/R_{t\star}\right)^2 \label{eq-tth-flux}
\end{equation}
where $F_t$ is the flux of the star with the transiting planet, $R_p$ is the 
radius of the planet, and $R_{t\star}$ is the radius of the star being 
transited. Substituting into equation~(\ref{eq-single-flux}), the generalized 
transit depth equation (in the absence of limb darkening or star spots) becomes 
\begin{equation}
\delta_o =
\left(\frac{F_t}{F_{total}}\right)\left(\frac{R_p}{R_{t\star}}\right)^2.
\label{eq-multi-radii}
\end{equation}

For a single star, $F_{total} = F_t$ and the transit depth expression simplifies 
to just the square of the size ratio between the planet and the star. However, 
for a multiple star system, the relationship between the observed transit depth 
and the true planetary radius depends upon the brightness ratio of the transited 
star to the total brightness of the system {\em and} on the stellar radius which 
changes depending on which star the planet is transiting: 
\begin{equation}
R_p = R_{t\star}\sqrt{\delta_o\frac{F_{total}}{F_t}}.\label{eq-radius-depth}
\end{equation}

The Kepler planetary candidates parameters are estimated assuming the star is a 
single star \citep{batalha11,burke14,mullally15}, and, therefore, may 
incorrectly report the planet radius if the stellar host is really a multiple 
star system.  The extra flux contributed by the companion stars will dilute the 
observed transit depth, and the derived planet radius depends on the size of 
star presumed to be transited. The ratio of the true planet radius, $R_p(true)$, 
to the observed planet radius assuming a single star with no companions, 
$R_p(observed)$, can be described as: 
\begin{equation}
X_R \equiv \frac{R_p(true)}{R_p(observed)} = 
\left(\frac{R_{t\star}}{R_{1\star}} 
\right)\sqrt{\frac{F_{total}}{F_t}}\label{eq-full-ratio},
\end{equation}
where $R_{1\star}$ is the radius of the (assumed single) primary star, and 
$F_t$ and $R_{t\star}$ are the brightness and the radius, respectively, of the 
star being transited by the planet.

This ratio reduces to unity in the case of a single star ($R_{t\star} \equiv 
R_{1\star}$ and $F_{total} \equiv F_t$).  For a multiple star system where the 
planet orbits the primary star ($R_{t\star} \equiv R_{1\star}$), the planet size 
is underestimated only by the flux dilution factor:
\begin{equation}
X_R \equiv \frac{R_p(true)}{R_p(observed)} = \sqrt{\frac{F_{total}}{F_1} 
}\label{eq-primary-ratio}.
\end{equation}
However, if the planet orbits one of the companion stars and not the primary 
star, then the ratio of the primary star radius ($R_{1\star}$) to the radius of 
the companion star being transited ($R_{t\star}$) affects the observed 
planetary 
radius, in addition to the flux dilution factor. 

\begin{figure}[t]
\includegraphics[angle=0,scale=0.5,keepaspectratio=true]{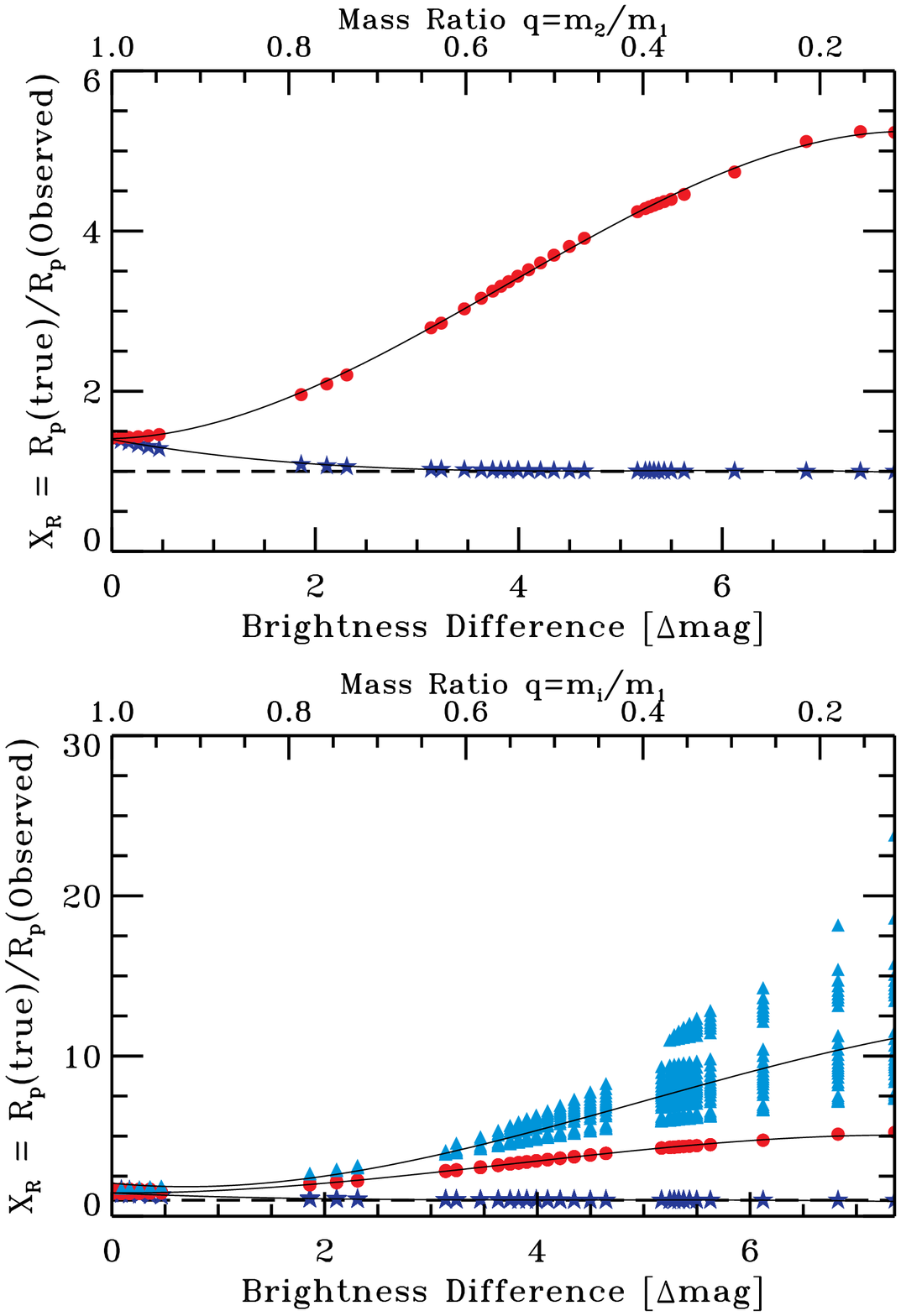} 
\caption{Radii corrections factors ($X_R$) are plotted as a function of 
companion-to-primary brightness ratios ({\it bottom axis}) and mass ratios ({\it 
top axis}) for possible binary systems ({\it top plot}) or triple ({\it bottom 
plot}) systems.  This figure is an example for the G-dwarf KOI-299; similar 
calculations have been made for every KOI. In each plot, the dark blue stars 
represent the correction factors if the planet orbits the primary star (equation 
\ref{eq-primary-ratio}); the red circles represent the correction factors if the 
planet orbits the secondary star, and the light blue triangles represent the 
correction factors if the planet orbits the tertiary star (equation 
\ref{eq-full-ratio}).  The lines are third order polynomials fit to the 
distributions.  Unity is marked with a horizontal dashed line.} 
\label{fig-koi299} 
\end{figure}
\begin{figure}[ht]
\includegraphics[angle=0,scale=0.5,keepaspectratio=true]{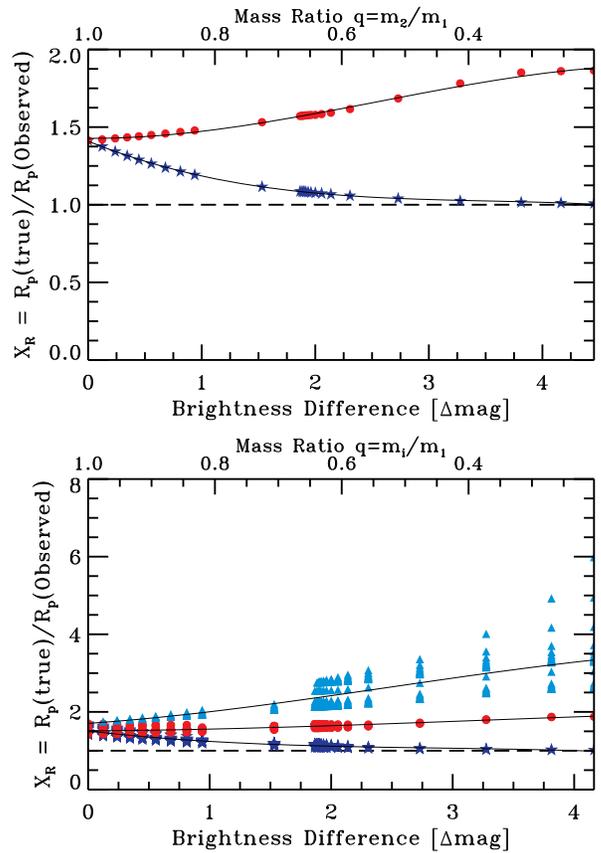}
\caption{This figure is the same as Figure \ref{fig-koi299}, but for the 
M-dwarf KOI-1085, demonstrating that the details of the derived correction 
factors are dependent upon the KOI properties.} 
\label{fig-koi1085}
\end{figure}

\section{Possible Companions from Isochrones}\label{sec-companions}
To explore the possible effects of the undetected stellar companions on the 
derived planetary parameters, we first assess what companions are possible for 
each KOI. For this work, we have downloaded the Cumulative Kepler candidate list 
and stellar parameters table from the NASA Exoplanet Archive.  The cumulative 
list is updated with each new release of the KOI lists\footnote{The 2014 October 
23 update to the cumulative KOI table was used in the work presented here: 
http://exoplanetarchive.ipac.caltech.edu/}; as a result, the details of any one 
star and planet may have changed since the analysis for this paper was done.  
However, the overall results of the paper presented here should remain largely 
unchanged. 

For the KOI lists, the stellar parameters for each KOI were determined by 
fitting photometric colors and spectroscopically derived parameters (where 
available) to the Dartmouth Stellar Evolution Database \citep{dotter08,huber14}. 
The planet parameters were then derived based upon the transit curve fitting and 
the associated stellar parameters.  Other stars listed in the Kepler Input 
Catalog or UKIRT imaging that may be blended with the KOI host stars were 
accounted for in the transit fitting, but, in general, as mentioned above, each 
planetary host star was assumed to be a single star.

We have restricted the range of possible bound stellar companions to each KOI 
host star by utilizing the same Dartmouth isochrones used to determine the 
stellar parameters.  Possible gravitationally bound companions are assumed to 
lie along on the same isochrone as the primary star.  For each KOI host star, we 
found the single best fit isochrone (characterized by mass, metallicity, and 
age) by minimizing the chi-square fit to the stellar parameters (effective 
temperature, surface gravity, radius, and metallicity) listed in the KOI table. 
 
We did not try to re-derive stellar parameters or independently find the best 
isochrone fit for the star; we simply identified the appropriate Dartmouth 
isochrone as used in the determination of the stellar parameters 
\citep{huber14}.  We note that there exists an additional uncertainty based upon 
the isochrone finding.  In this work, we did not try to re-derive the stellar 
parameters of the host stars, but rather, we simply find the appropriate 
isochrone that matches the KOI stellar parameters.  Thus, any errors in the 
stellar parameters derivations in the KOI list are propagated here.  This is 
likely only a significant source of uncertainty for nearly equal brightness 
companions.  

Once an isochrone was identified for a given star, all stars along an isochrone 
with (absolute) Kepler magnitudes fainter than the (absolute) Kepler magnitude 
of the host star were considered to be viable companions; i.e., the primary host 
star was assumed to be the brightest star in the system.  The fainter companions 
listed within that particular isochrone were then used to establish the range of 
possible planetary radii corrections (equation \ref{eq-full-ratio}) assuming the 
host star is actually a binary or triple star.  Higher order (e.g., quadruple) 
stellar multiples are not considered here as they represent only $\sim3$\% of 
the stellar population \citep{raghavan10}. 

\begin{figure*}[t]
\includegraphics[angle=90,scale=0.75,keepaspectratio=true]{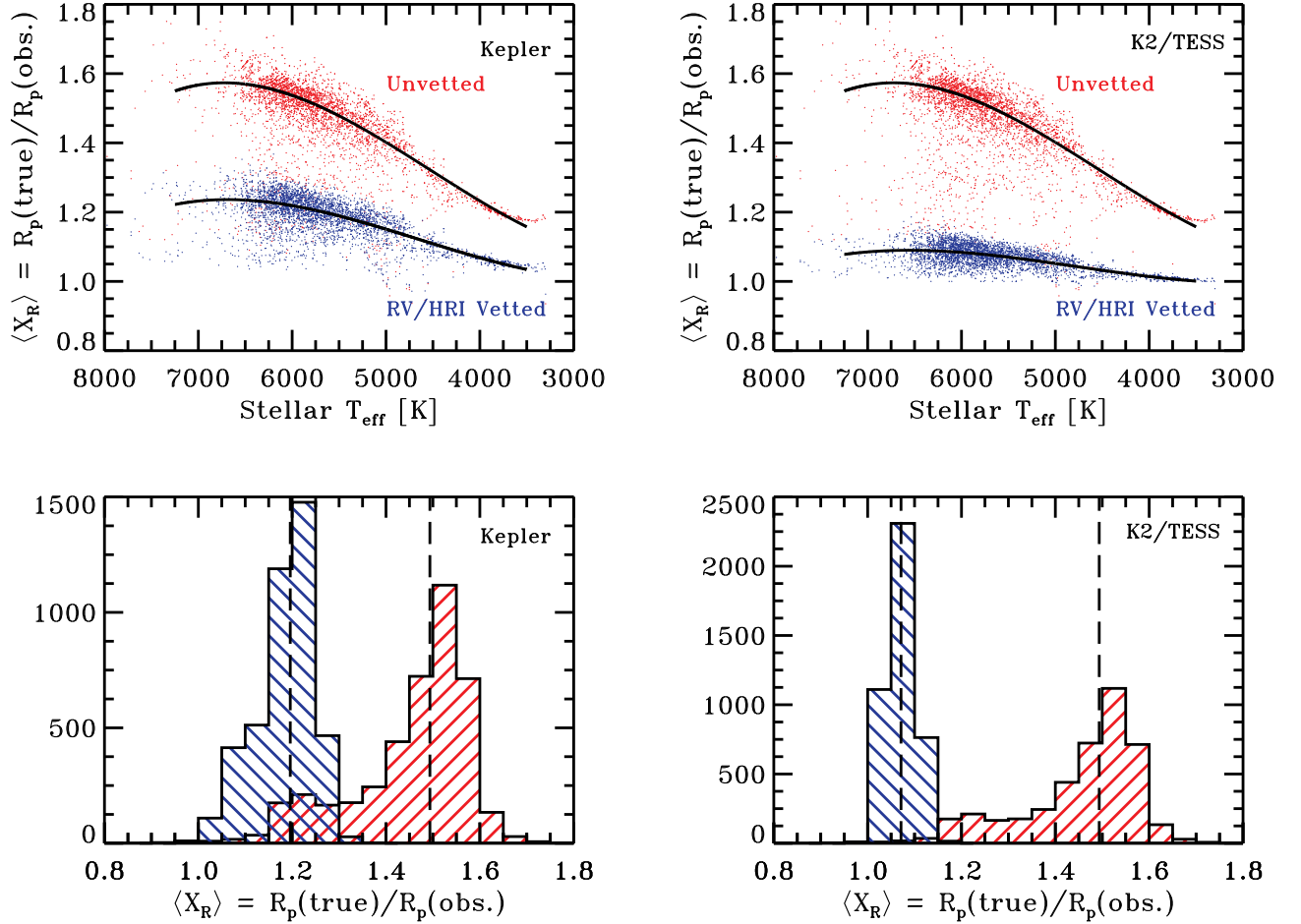} 
\caption{{\it Top:} The mean correction factor $\langle X_R \rangle$ for each 
KOI is displayed as a function of the effective temperature of the primary star 
(see section~\ref{sec-X}). The black curves are $3^{rd}$-order polynomials 
fitted to the distributions (equation \ref{eq-factor-unvetted}). {\it Bottom:} 
Histograms of the correction factors displayed in the top panels.  The vertical 
dashed lines mark the medians of the distributions. {\it Left:} The correction 
factors are computed for the Kepler Cumulative Kepler Objects of Interest list. 
{\it Right:} The corrections are computed for the KOIs but assuming the KOIs are 
10 times closer as may be the case for K2 and TESS. The red points and 
histograms assume each KOI is single as is the case for the published KOI list; 
the blue points and histograms assume that each KOI has been vetted with radial 
velocity (RV) and high resolution imaging (HRI), and all stellar companions with 
orbital periods of 2 years or shorter and all stellar companions located at 
angular distances of $\geq 0.1^{\prime\prime}$ have been detected and accounted 
for in the planetary radii determinations (see Section \ref{sub-vetting}).  For 
the vetted stars, the correction factors are only for undetected stellar 
companions; detected companions have been assumed to be accounted for in the 
planet radii determinations.} 
\label{fig-mean-factor}
\end{figure*}

We have considered six specific multiplicity scenarios:
\begin{enumerate}
\item single star ($X_r \equiv 1.0$)
\item binary star -- planet orbits primary star
\item binary star -- planet orbits secondary star
\item triple star -- planet orbits primary star
\item triple star -- planet orbits secondary star
\item triple star -- planet orbits tertiary star.
\end{enumerate}
Based upon the brightness and size differences between the primary star and the 
putative secondary or tertiary companions, we have calculated for each KOI the 
possible factor by which the planetary radii are underestimated ($X_R$). If the 
star is single, the correction factor is unity, and if, in a multiple star 
system, the planet orbits the primary star, only flux dilution affects the 
observed transit depth and the derived planetary radius 
(eq.~\ref{eq-primary-ratio}). 

For the scenarios where the planets orbits the secondary or tertiary star, the 
planet size correction factors (eq.~\ref{eq-full-ratio}) were determined only 
for stellar companions where the stellar companion could physically account for 
the observed transit depth.  If more than 100\% of the stellar companion light 
had to be eclipsed in order to produce the observed transit in the presence of 
the flux dilution, then that star (and all subsequent stars on the isochrone 
with lower mass) was not considered viable as a potential source of the transit. 
 For example, for an observed 1\% transit, no binary companions can be fainter 
than the primary star by 5 magnitudes or more; an eclipse of such a secondary 
star would need to be more than 100\% deep.  The stellar brightness limits were 
calculated independently for each planet within a KOI system so as to not assume 
that all planets within a system necessarily orbited the same star.

Figures \ref{fig-koi299} and \ref{fig-koi1085} show representative correction 
factors ($X_R$) for KOI-299 (a G-dwarf with a super-Earth sized $R_p = 1.8\pm 
0.24R_\oplus$ planet) and for KOI-1085 (an M-dwarf with an Earth-sized $R_p = 
0.92\pm0.13 R_\oplus$ planet). The planet radius correction factors ($X_R$) are 
shown as a function of the companion--to--primary brightness ratio (bottom 
x-axis of plots) and the companion--to--primary mass ratio (top x-axis of plots) 
and are determined for the KOI assuming it is a binary-star system (top plot) or 
a triple-star system (bottom plot).  

The amplitude of the correction factor ($X_R$) varies strongly depending on the 
particular system and which star the planet may orbit.  If the planet orbits the 
primary star, then the largest the correction factors are for equal brightness 
companions ($\sqrt{2} \sim 1.4$ for a binary system and $\sqrt{3} \sim 1.7$ for 
a triple system) with an asymptotic approach to unity as the companion stars 
become fainter and fainter. If the planet orbits the secondary or tertiary star, 
the planet radius correction factor can be significantly larger -- ranging from 
$X_R \gtrsim 2 - 5$ for binary systems and $X_R \gtrsim 2 - 20$ or more for 
triple systems -- depending on the size and brightness of the secondary or 
tertiary star.

\section{Mean Radii Correction Factors ($X_R$)}\label{sec-X}

It is important to recognize the full range of the possible correction factors, 
but in order to have a better understanding of the statistical correction any 
given KOI (or the KOI list as a whole) may need, we must understand the mean 
correction for any one multiplicity scenario and convert these into a single 
mean correction factor for each star. To do this, we must take into account the 
probability the star may be a multiple star, the distribution of mass ratios if 
the star is a multiple, the probability that the planet orbits any one star if 
the stellar system has multiple stars, and whether or not the star has been 
vetted (and how well it is has been vetted) for stellar companions.  

In order to calculate an average correction factor for each multiplicity 
scenario, we have fitted the individual scenario correction factors as a 
function of mass ratio with a 3$^{rd}$-order polynomial (see Fig. 
\ref{fig-koi299} and \ref{fig-koi1085}). Because the isochrones are not evenly 
sampled in mass, taking a mean straight from the isochrone points would skew the 
results; the polynomial parameterization of the correction factor as a function 
of the mass ratio enables a more robust determination of the mean correction 
factor for each multiplicity scenario.

If the companion--to--primary mass ratio distribution was uniform across all 
mass ratios, then a straight mean of the correction values determined from each 
polynomial curve would yield the average correction factor for each multiplicity 
scenario.  However, the mass ratio distribution is likely not uniform, and we 
have adopted the form displayed in Figure 16 of \citet{raghavan10}. That 
distribution is a nearly-flat frequency distribution across all mass ratios with 
a $\sim2.5\times$ enhancement for nearly equal mass companion stars ($q \gtrsim 
0.95$).   This distribution is in contrast to the Gaussian distribution shown in 
\citet{dm91}; however, the more recent results of \citet{raghavan10} incorporate 
more stars, a broader breadth of stellar properties, and multiple companion 
detection techniques.    

The mass ratio distribution is convolved with the polynomial curves fitted for 
each multiplicity scenario, and a weighted mean for each multiplicity scenario 
was calculated for every KOI.  For example, in the case of KOI-299 (Fig. 
\ref{fig-koi299}),  the single star mean correction factor is 1.0 (by 
definition). For the binary star cases, the average scenario correction factors 
are 1.14 (planet orbits primary) and 2.28 (planet orbits secondary); for the 
triple stars cases, the correction factors are 1.16 (planet orbits primary), 
2.75 (planet orbits secondary), and 4.61 (planet orbits tertiary).  For KOI-1085 
(Fig. \ref{fig-koi1085}), the weighted mean correction factors are 1.18, 1.56, 
1.24, 1.61, and 2.29, respectively.

To turn these individual scenario correction factors into an overall single mean 
correction factor $\langle X_R \rangle$ per KOI, the six scenario corrections 
are convolved with the probability that a KOI will be a single star, a binary 
star, or a triple star. The multiplicity rate of the Kepler stars is still 
unclear \citep{wang14b}, and, indeed, there may be some contradictory evidence 
for the the exact value for the multiplicity rates of the KOI host stars 
\citep[e.g.,][]{horch14,wang14b}, but the multiplicity rates appear to be near 
$40\%$, similar to the general field population. In the absence of a more 
definitive estimate, we have chosen to utilize the multiplicity fractions from 
\citet{raghavan10}: a 54\% single star fraction, a 34\% binary star fraction, 
and a 12\% triple star fraction \citep{raghavan10}. We have grouped all higher 
order multiples ($3+$) into the single category of ``triples'', given the 
relatively rarity of the quadruple and higher order stellar systems.   For the 
scenarios where there are multiple stars in a system, we have assumed that the 
planets are equally likely to orbit any one of the stars (50\% for binaries, 
33.3\% for triples).

The final mean correction factors $\langle X_R \rangle$ per KOI are displayed in 
Figure~\ref{fig-mean-factor}; the median value of the correction factor and the 
dispersion around that median is $\langle X_R \rangle = 1.49 \pm 0.12$.  This 
median correction factor implies that assuming a star in the KOI list is single, 
in the absence of any (observational) companion vetting, yields a statistical 
bias on the derived planetary radii where the radii are underestimated, {\it on 
average}, by a factor $\sim 1.5$, and the mass density of the planets are 
overestimated by a factor of $\sim 1.5^3 \sim 3$.

From Figure \ref{fig-mean-factor}, it is clear that the mean correction factor 
$\langle X_R \rangle$ depends upon the stellar temperature of the host star.  As 
most of the stars in the KOI list are dwarfs, the lower temperature stars are 
typically lower mass stars and, thus, have a smaller range of possible stellar 
companions.  Thus, an average value for the correction factor 1.5 represents the 
sample as a whole, but a more accurate value for the correction factor can be 
derived for a given star, with a temperature between $3500 \lesssim T_{eff} 
\lesssim 7500$K, using the fitted $3^{rd}$-order polynomial:
\begin{equation}
\langle X_R \rangle = a_3(T_{eff})^3 + a_2(T_{eff})^2 + a_1(T_{eff}) + a_0
\label{eq-factor-unvetted}
\end{equation}
where $a_3 = -1.19118\times10^{-11}, a_2 = 1.61749\times10^{-7}, a_1 =  
-0.000560, {\ \rm and\ } a_0 = 1.64668$.

In the absence of any specific knowledge of the stellar properties (other than 
the effective temperature) and in the absence of any radial velocity or high 
resolution imaging to assess the specific companion properties of a given KOI, 
(see section \ref{sub-vetting}), the above parameterization (equation 
\ref{eq-factor-unvetted}) can be used to derive a mean radii correction factor 
$\langle X_R \rangle$ for a given star.  For G-dwarfs and hotter stars, the 
correction factor is near $\langle X_R \rangle \sim 1.6$.  As the stellar 
temperature (mass) of the primary decreases to the range of M-dwarfs, the 
correction factor can be as low as $\langle X_R \rangle \sim 1.2$.

\begin{figure}[t]
\includegraphics[angle=0,scale=0.5,keepaspectratio=true]{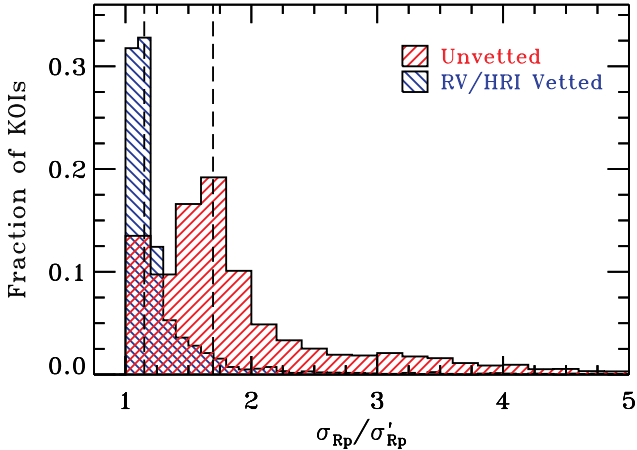} 
\caption{The distribution of the ratio of the total planetary radii 
uncertainties ($\sigma_{R_p}$) to the quoted radii uncertainties 
($\sigma^\prime_{R_p}$)from the cumulative KOI list (see equation 
\ref{eq-total-unc}).  For the red histogram, it is assumed that the KOIs are 
single as is the case in the published KOI list; for the blue histogram, it is 
assumed that each KOI has been vetted with radial velocity (RV) and high 
resolution imaging (see Section \ref{sub-vetting}).  The vertical dashed lines 
represent the median values of the distributions: $\langle 
\sigma_{R_p}/\sigma^{\prime}_{R_p} \rangle = 1.70$ for the unvetted KOIs and 
$\langle \sigma_{R_p}/\sigma^{\prime}_{R_p} \rangle = 1.15$ for the vetted KOIs 
(see section \ref{sub-vetting}).} 
\label{fig-ratio-unc}
\end{figure}

\subsection{Planet Radius Uncertainty Term from $\langle 
X_R\rangle$}\label{sub-unc}

The mean correction factor is useful for understanding how strongly the 
planetary radii may be underestimated, but an additional uncertainty term 
derived from the mean radius correction factor is potentially more useful as it 
can be added in quadrature to the formal planetary radii uncertainties.  The 
formal uncertainties, presented in the KOI list, are derived from the 
uncertainties in the transit fitting and the uncertainty in the knowledge of the 
stellar radius, and they are calculated assuming the KOIs are single stars.  We 
can estimate an additional planet radius uncertainty term based upon the mean 
radii correction factor as
\begin{equation}
\sigma_{X_R} = |\langle X_R\rangle R_p - R_p| = |\langle X_R\rangle - 1.0|R_p 
\label{eq-factor-unc}
\end{equation}
where $R_p$ is the observed radius of the planet. Adding in quadrature to the 
reported uncertainty, a more complete uncertainty on the planetary radius can be 
reported as
\begin{equation}
\sigma_{R_p} = \left( (\sigma^{\prime}_{R_p})^2 + (\sigma_{X_R})^2   
\right)^{1/2}\label{eq-total-unc}
\end{equation}
where $\sigma^{\prime}_{R_p}$ is the uncertainty of the planetary radius as 
presented in the KOI list.

The distribution of the ratio of the more complete KOI radius uncertainties 
($\sigma_{R_p}$) to the reported KOI radius uncertainties 
($\sigma^{\prime}_{R_p}$) is shown in Figure~\ref{fig-ratio-unc}.  Including the 
possibility that a KOI may be a multiple star increases the planetary radii 
uncertainties. While the distribution has a long tail dependent upon the 
specific system, the planetary radii uncertainties are underestimated as 
reported in the KOI list, {\it on average}, by a factor of 1.7. 

\begin{figure*}[t]
\includegraphics[angle=90,scale=0.75,keepaspectratio=true]{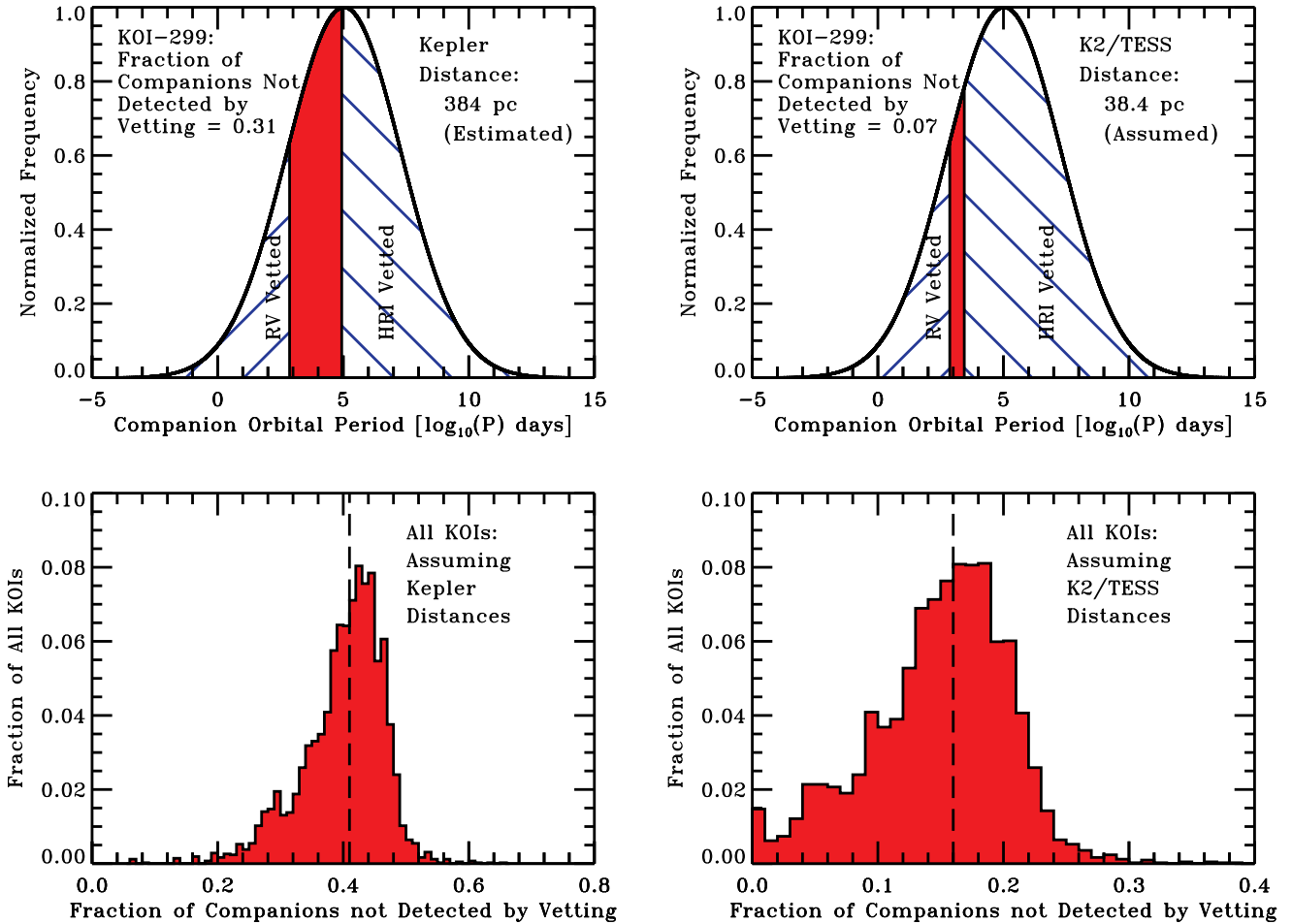}
\caption{{\it Top Left:} Example of the stellar companion period distribution 
that is vetted by radial velocity (RV) monitoring and  high resolution imaging 
(HRI) for KOI-299.  The Gaussian curve, normalized to unity, is the log-normal 
orbital period distribution of stellar companions \citep{raghavan10}, and the 
hatched regions mark where the potential observational vetting is assumed to 
have detected all companions in that period range. The solid (red) region of the 
Gaussian corresponds to the fraction of companions (31\% for KOI-299) that would 
remain undetected by the RV and HRI observations (see section 
\ref{sub-vetting}). {\it Bottom Left:} The distribution of the fraction of 
companions across all KOIs that remain undetected by observational radial 
velocity and high resolution imaging vetting (i.e., the red area in the top 
figure but for all KOIs). The vertical dashed line represents the median 
fraction ($41\%$) of companions left undetected by the observational vetting of 
all KOIs. {\it Top Right:} The same as the {\it top left} figure but assuming 
that KOI-299 is 10 times closer as may be the case for K2/TESS targets. At such 
a close distance for a hypothetical KOI-299, RV and HRI vetting leaves only 
about 7\% of the possible stellar companions undetected. {\it Bottom Right:} As 
the {\it bottom left} figure but assuming all of the KOI host stars are 10 times 
closer as may be the case for K2/TESS. The vertical dashed line represents the 
median fraction ($16\%$) of companions left undetected if the KOIs were at these 
closer distances.} 
\label{fig-logP} 
\end{figure*}

\subsection{Effectiveness of Companion Vetting}\label{sub-vetting}

The above analysis has assumed that the KOIs have undergone no companion 
vetting, as is the assumption in the current KOI list.  In reality, the Kepler 
Project has funded a substantial ground-based follow-up observation program 
which includes radial velocity vetting and high resolution imaging.  In this 
section, we explore the effectiveness of the observational vetting.

The observational vetting reduces the fraction of undetected companions.  If 
there is no vetting or all stars are assumed to be single, as is the case for 
the published KOI list, then the fraction of undetected companions is 100\% and 
the mean correction factors $\langle X_R \rangle$ are as presented above.  If 
every stellar companion is detected and accounted for in the planetary parameter 
derivations, then the fraction of undetected companions is 0\%, and the mean 
correction factors are unity. Reality is somewhere in between these two 
extremes. 

To explore the effectiveness of the observational vetting on reducing the radii 
corrections factors (and the associated radii uncertainties), we have assumed 
that every KOI has been vetted equally, and all companions within the reach of 
the observations have been detected and accounted. Thus, the corrections factors 
depend only on the fraction of companions stars that remain out of the reach of 
vetting and undetected.

In this simulation, we have assumed that all companions with orbital periods of 
2 years or less and all companions with angular separations of $0.1\arcsec$ or 
greater have been detected.  This, of course, will not quite be true as random 
orbital phase effects, inclination effects, companion mass distribution, stellar 
rotation effects, etc. will diminish the efficiency of the observations to 
detect companions.  We recognize the simplicity of these assumptions; however, 
the purpose of this section is to assess the usefulness of observational vetting 
on reducing the uncertainties of the planetary radii estimates, not to explore 
fully the sensitivities and completeness of the vetting.  

Typical follow-up observations include stellar spectroscopy, a few radial 
velocity measurements, and high resolution imaging.  The radial velocity 
observations usually include $2-3$ measurements over the span of $6-9$ months 
and are typically sufficient to identify potential stellar companions with 
orbital periods of $\lesssim 1-2$ years or less. While determining full orbits 
and stellar masses for any stellar companions detected typically requires more 
intensive observing,  we have estimated that 3 measurements spanning $6 - 9$ 
months is sufficient to enable the detection of an RV trend for orbital periods 
of $\sim 2$ years or less and mark the star as needing more detailed 
observations.  The amplitude of the RV signature, and hence the ability to 
detect companions, does depend upon the masses of the primary and companion 
stars; massive stars with low mass companions will display relatively low RV 
signatures. However, RV vetting for the Kepler program has been done at a level 
of $\lesssim 100 - 200$ m/s, which is sufficient to detect (at $\gtrsim 
4-5\sigma$) a late-type M-dwarf companion in a two-year orbit around a mid 
B-dwarf primary.  Indeed, the RV vetting is made even more effective by 
searching for companions via spectral signatures \citep{kolbl15}.

The high resolution imaging via adaptive optics, ``lucky imaging'', and/or 
speckle observations typically has resolutions of $0.02\arcsec - 0.1\arcsec$ 
\citep[e.g.,][]{howell11,horch12,lb12, 
adams12,adams13,dressing14,law14,lb14,wang14a, everett14, horch14, gilliland14, 
star14}.  Based upon Monte Carlo simulations in which we have averaged over 
random orbital inclinations and eccentricities, we have calculated the fraction 
of time within its orbit a companion will be detectable via high resolution 
imaging.  With typical high resolution imaging of 0.05\arcsec, we have estimated 
that $\gtrsim 50\%$ of the stellar companions will be detected at one full-width 
half-maximum (FWHM=0.05\arcsec) of the image resolution and beyond and $>90$\% 
at $\gtrsim2$ FWHM (0.1\arcsec) of the image resolution and beyond.  

To determine what fraction of possible stellar companions would be detected in 
such a scenario, we have used the nearly log-normal orbital period distribution 
from \citet{raghavan10}.  To convert the high resolution imaging limits into 
period-limits, we have estimated the distance to each KOI by determining a 
distance modulus from the observed Kepler magnitude and the absolute Kepler 
magnitude associated with the fitted isochrone.  The median distance to the KOIs 
was found to be $\sim 900$ pc, corresponding to $\sim 90$ AU for 0.1\arcsec\ 
imaging.  Using the isochrone stellar mass, the semi-major axis detection limits 
were converted to orbital period limits (assuming circular orbits).  

Combining the 2-year radial velocity limit and the $0.1\arcsec$ imaging limit, 
we were able to estimate the fraction of undetected companions for each 
individual KOI (see Figure~\ref{fig-logP}).   The distribution of the fraction 
of undetected companions ranges from $\sim 20 - 60$\% and, on average, the 
ground-based observations leave $\approx 41\%$ of the possible companions 
undetected for the KOIs (see Figure~\ref{fig-logP}).  

The mean correction factors $\langle X_R \rangle$ are only applicable to the 
undetected companions.  For the stars that are vetted with radial velocity 
and/or high resolution imaging, the intrinsic stellar companion rate for the 
KOIs of 46\% \citep{raghavan10} is reduced by the unvetted companion fraction 
for each KOI. That is, we assume that companion stars detected in the vetting 
have been accounted for in the planetary radii determinations, and the unvetted 
companion fraction is the relevant companion rate for determining the correction 
factors. In the KOI-299 example (Fig.~\ref{fig-logP}), the undetected companion 
rate used to calculate the mean radii correction factor is $0.46 \times 0.31 = 
0.1426$. This lower fraction of undetected companions in turn reduces the mean 
correction factors for the vetted stars which are displayed in 
Figure~\ref{fig-mean-factor} (blue points).

Instead of a mean correction factor of $\langle X_R \rangle \sim 1.5$, the 
average correction factor is $\langle X_R \rangle =1.20 \pm 0.06$ if the stars 
are vetted with radial velocity and high resolution imaging.  The mean 
correction factor still changes as a function of the primary star effective 
temperature but the dependence is much more shallow with coefficients for 
equation~\ref{eq-factor-unvetted} of $a_3 = -6.73847\times10^{-12}, a_2 = 
9.38966\times10^{-8}, a_1 = -0.000352, {\ \rm and\ } a_0 = 1.40391$ (see 
Figure~\ref{fig-mean-factor}).

\begin{figure*}[t]
  \includegraphics[angle=90,scale=0.75,keepaspectratio=true]{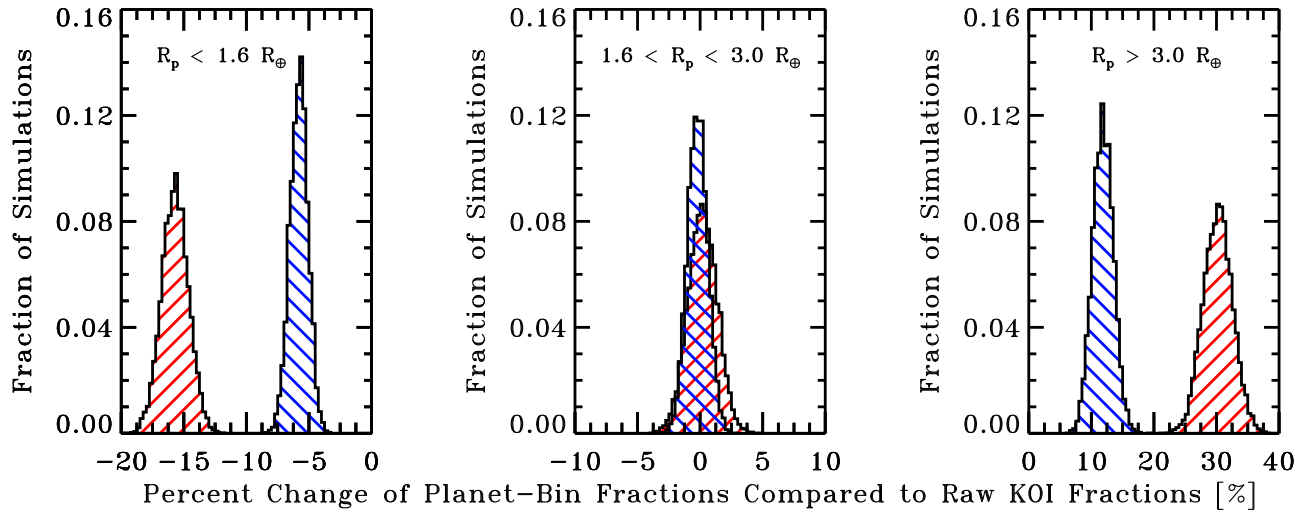}
  \caption{The percent change of the measured occurrence rates from the raw KOI 
list (i.e., uncorrected for completeness) caused by undetected stellar 
companions and assuming all of the stars in the published KOI list are single. 
The fractional changes are computed for three different planet size bins as 
labeled in the plots.    The results of assuming all the stars are single (i.e., 
unvetted) are in the red histogram and the results of vetting the all of the 
stars with radial velocity and high resolution imaging (section 
\ref{sub-vetting}) are shown in the blue histogram.  The histograms are based 
upon Monte Carlo simulations described in section \ref{sec-rates}} 
\label{fig-occurrence-rates}
\end{figure*}

\subsection{K2 and TESS}\label{sub-k2tess}
The above analysis has concentrated on the Kepler Mission and the associated KOI 
list, but the same effects will apply to all transit surveys including K2 
\citep{howell14} and TESS \citep{ricker14}.  If the planetary host stars from 
K2 and TESS are also assumed to be single with no observational vetting, the 
planetary radii will be underestimated by the same amount as the Kepler KOIs 
(Fig.~\ref{fig-mean-factor} and eq.~\ref{eq-factor-unvetted}).

Many K2 targets and nearly all of TESS targets will be stars that are typically 
$4-5$ magnitudes brighter than the stars observed by Kepler, and therefore, 
K2 and TESS targets will be $\sim 10$ times closer than the Kepler targets.  
The effectiveness of the radial velocity vetting will remain mostly unaffected 
by the brighter and closer stars, but the effectiveness of the high resolution 
imaging will be significantly enhanced.  Instead of probing the stars to within 
$\sim 100$AU, the imaging will be able to detect companion stars within $\sim 
10$ AU of the stars.  

As a result, the fraction of undetected companions will decrease significantly. 
Even for the Kepler stars that undergo vetting via radial velocity and high 
resolution imaging, $\sim 40$\% of the companions remain undetected. But for the 
stars that are 10 times closer that fraction decreases to $\sim 16$\% (see 
Figure~\ref{fig-logP}).  This has the strong benefit of greatly reducing the 
mean correction factors for the stars that are observed by K2 and TESS and are 
vetted for companions with radial velocity and high resolution imaging.

The mean correction factor for vetted K2/TESS-like stars is only $\langle X_R 
\rangle = 1.07 \pm 0.03$.  The correction factor has a much flatter dependence 
on the primary star effective temperature, because the majority of the possible 
stellar companions are detected by the vetting.  The coefficients for equation 
\ref{eq-factor-unvetted} become $a_3 = -4.12309\times10^{-12}, a_2 = 
5.89709\times10^{-8}, a_1 = -0.000242, {\ \rm and\ } a_0 = 1.30060$.   The mean 
radii correction factors for vetted K2/TESS planetary host stars correspond to a 
correction to the planetary radii uncertainties of only $\sim 2$\%, in 
comparison to a correction of $\sim 70$\% if the K2/TESS stars remain unvetted. 

For K2 and TESS, where the number of candidate planetary systems may outnumber 
the KOIs by an order of magnitude (or more), single epoch high resolution 
imaging may prove to be the most important observational vetting performed.  
While the imaging will not reach the innermost stellar companions, radial 
velocity observations require multiple visits over a baseline comparable to the 
orbital periods an observer is trying to sample.  In contrast, the high 
resolution imaging requires a single visit (or perhaps one per filter on a 
single night) and will sample the majority of the expected stellar companion 
period distribution.

\section{Effect of Undetected Companions on the Derived Occurrence 
Rates}\label{sec-rates}

Understanding the occurrence rates of the Earth-sized planets is one of the 
primary goals of the Kepler mission and one of the uses of the KOI list 
\citep{borucki10}. It has been shown that the transition from ''rocky'' to 
''non-rocky'' planets occurs near a radius of $R_p = 1.6\ R_\oplus$ and the 
transition is very sharp \citep{rogers14}.   However, the amplitude of the 
uncertainties resulting from undetected companions may be large enough to push 
planets across this boundary and affect our knowledge of the fraction of 
Earth-sized planets.  

We have explored the possible effects of undetected companions on the derived 
occurrence rates. The planetary radii can not simply be multiplied by a mean 
correction factor $X_R$, as that factor is only a measure of the statistical 
uncertainty of the planetary radius resulting from assuming the stars are single 
and only a fraction of the stars are truly multiples.  Instead a Monte Carlo 
simulation has been performed to assign randomly the effect of unseen companions 
on the KOIs.

The simulation was performed 10,000 times for each KOI.  For each realization of 
the simulation, we have randomly assigned the star to be single, binary, or 
triple star via the 54\%, the 34\% and the 12\% fractions \citep{raghavan10}.  
If the KOI is assigned to be a single star, the mean correction factors for the 
planets in that system are unity: $\langle X_R \rangle = 1$.  If the KOI star is 
a multiple star system, we have randomly assigned the stellar companion masses  
according to the masses available from the fitted isochrones and using the mass 
ratio distribution of \citep{raghavan10}.  Finally, the planets are randomly 
assigned to the primary or to the companion stars (i.e., 50\% fractions for 
binary stars and 33.3\% fractions for triple stars).  Once the details for the 
system are set for a particular realization, the final correction factor for the 
planets are determined from the polynomial fits for the individual multiplicity 
scenarios (e.g., Fig.~\ref{fig-koi299} and \ref{fig-koi1085}).  

For each set of the simulations, we compiled the fraction of planets within the 
following planet-radii bins: $R_p\leq 1.6\ R_\oplus$; $1.6 < R_p \leq 3.0\ R_p$; 
$3.0 < R_p \leq 10 \ R_\oplus$ corresponding to Earth-sized, 
super-Earth/mini-Neptune-sized, and Neptune-to-Jupiter-sized planets. The raw 
fractions directly from the KOI-list, for these three categories of planets, are 
33.3\%, 46.0\%, and 20.7\%. Note that these are the raw fractions and are not 
corrected for completeness or detectability as must be done for a true 
occurrence rate calculation; these fractions are necessary for comparing how 
unseen companions affect the determination of fractions. Finally, we repeated 
the simulations, but using the undetected multiple star fractions after vetting 
with radial velocity and high resolution imaging had been performed, thus, 
effectively increasing the fraction of stars with correction factors of unity.

The distributions of the change in the fractions of planets in each planet 
category, compared to the raw KOI fractions, are shown in 
Figure~\ref{fig-occurrence-rates}.  If the occurrence rates utilize the 
assumed-single KOI list (i.e., unvetted), then the Earth-sized planet fraction 
may be overestimated by as much as $15-20$\% and the giant-planet fraction may 
be underestimated by as much as 30\%.  Interestingly, the fraction of 
super-Earth/mini-Neptune planets does not change substantially; this is a result 
of smaller planets moving into this bin, and larger planets moving out of the 
bin.  In contrast, if all of the KOIs undergo vetting via radial velocity and 
high resolution imaging, the fractional changes to these bin fractions are much 
smaller: $5-7$\% for the Earth-sized planets and $10-12$\% for the 
Neptune/Jupiter-sizes planets.

\section{Summary}\label{sec-summary}
We present an exploration of the effect of undetected companions on the measured 
radii of planets in the Kepler sample.  We find that if stars are assumed to be 
single (as they are in the current Kepler Objects of Interest list) and no 
companion vetting with radial velocity and/or high resolution imaging is 
performed, the planetary radii are underestimated, {\it on average}, by a factor 
of $\langle X_R \rangle = 1.5$, corresponding to an overestimation of the planet 
bulk density by a factor of $\sim 3$.  Because lower mass stars will have a 
smaller range of stellar companion masses than higher mass stars, the planet 
radius mean correction factor has been quantified as a function of stellar 
effective temperature.

If the KOIs are vetted with radial velocity observations and high resolution 
imaging, the planetary radius mean correction necessary to account for 
undetected companions is reduced significantly to a factor of $\langle X_R 
\rangle = 1.2$.  The benefit of radial velocity and imaging vetting is even 
more powerful for missions like K2 and TESS, where the targets are, on average, 
ten times closer than the Kepler Objects of Interest.  With vetting, the 
planetary radii for K2 and TESS targets will only be underestimated, on 
average, by 10\%.  Given the large number of candidates expected to be 
produced by K2 and TESS, single epoch high resolution imaging may be the most 
effective and efficient means of reducing the mean planetary radius correction 
factor.  

Finally, we explored the effects of undetected companions on the occurrence 
rate calculations for Earth-sized, super-Earth/mini-Neptune-sized, and 
Neptune-sized and larger planets.  We find that if the Kepler Objects of 
Interest are all assumed to be single (as they currently are in the KOI list), 
then the fraction of Earth-sized planets may be overestimated by as much as 
15-20\% and the fraction of large planets may be underestimated by as much as 
30\%

The particular radial velocity observations or high resolution imaging vetting 
that any one KOI may (or may not) have undergone differs from star to star. 
Companion vetting simulations presented here show that a full understanding and 
characterization of the planetary companions is dependent upon also 
understanding the presence of stellar companions, but is also dependent upon 
understanding the limits of those observations. For a final occurrence rate 
determination of Earth-sized planets and, more importantly, an uncertainty on 
that occurrence rate, the stellar companion detections (or lack thereof) must be 
taken into account.

\acknowledgments

The authors would like to thank Ji Wang, Tim Morton, and Gerard van Belle for 
useful discussions during the writing of this paper. This research has made use 
of the NASA Exoplanet Archive, which is operated by the California Institute of 
Technology, under contract with the National Aeronautics and Space 
Administration under the Exoplanet Exploration Program. Portions of this work 
were performed at the California Institute of Technology under contract with 
the National Aeronautics and Space Administration.

%\newpage

\end{document}